\documentclass[a4,11pt,reqno]{amsart}
\usepackage[english]{babel}    \usepackage[latin1]{inputenc}   \usepackage[T1]{fontenc}     \usepackage[french]{minitoc}
\usepackage[nice]{nicefrac}    \usepackage{latexsym,amsfonts}  
\usepackage{graphics}    \usepackage{ulem}       \usepackage{hhline}    \usepackage{dsfont}    \usepackage{mathrsfs}
\usepackage{fancyhdr}    \usepackage{amsmath}    \usepackage{amssymb}   \usepackage{rotating}  \usepackage{fancybox}
\usepackage{color}       \usepackage{colortbl}   \usepackage{setspace}  \usepackage{enumerate} \usepackage{amsthm}
\usepackage{multicol}    \usepackage{pifont}     \usepackage{amsthm}    \usepackage{varioref}  \usepackage{textcomp}
\usepackage{lmodern}     \usepackage{mathpazo}   \usepackage{euscript}  \usepackage[pdftex]{hyperref}
\numberwithin{equation}{section}  \makeatletter\@addtoreset{equation}{section}
   \DeclareMathSymbol{\subsetneqq}{\mathbin}{AMSb}{36}

\begin{document}

\title[]{\textsc{Integral Transforms Connecting the Hardy space with Barut-Girardello Spaces}}
\author{\textsc{Zouhair Mouayn}}

\date{}

\maketitle

\vspace*{-.6cm}
\begin{center}
\small{  Department of Mathematics, Faculty of Sciences and Technics (M'Ghila),\\
 Sultan Moulay Slimane University, BP 523, B\'{e}ni Mellal, Morocco \\
 (E-mail: mouayn@fstbm.ac.ma)}
\end{center}

\begin{abstract}
We construct a one parameter family of integral transforms connecting the
classical Hardy space with a class of weighted Bergman spaces called Barut-Girardello spaces.
\end{abstract}

\section{Introduction}

The paper deal with we the construction of a one-parameter family of
integral transforms that connect the classical Hardy space $\mathcal{H}%
_{+}^{2}(\Bbb{R}) $ of complex-valued square integrable
functions $f(x) $ on the real line, \ whose Fourier transform
are supported by the positive real semi-axis with  Barut-Girardello spaces (\cite[p.51]{1}) which are  weighted Bergman space, denoted $\frak{F}%
_{\sigma }(\Bbb{C}) ,$ consisting of analytic functions $\varphi
(z) $ on the complex plane $\Bbb{C}$ , that are square
integrable with respect to the measure $|z|^{2\sigma -1}K_{%
\frac{1}{2}-\sigma }(2|z|) d\lambda (z)
$, $K_{\nu }(.) $ is the MacDonald function and $d\lambda $
being the planar Lebesgue measure  and $\sigma =\frac{1}{2},1,\frac{3}{2}%
,2,\cdots$ is a parameter.

The essence of our method consists on a coherent states analysis. Precisely, we will exploit some known results in \cite[pp.59-62]{2}  to construct a class of coherent states of Barut-Girardello type which belong
to the Hardy space and solve its identity. Therefore, the associated
coherent states transform turns out to be the integral transform we are
concerned with.

In the next section we recall briefly a well known formalism of coherent
states with their corresponding coherent state transforms. In section 3,
some basic facts on Hardy spaces are reviewed. Section 4 is devoted to the
definition of the Barut-Girardello spaces. In Section 5, we establish an
integral transform linking the Hardy space with Barut-Girardello spaces.

\section{Coherent states}

Let us recall a well known general formalism (\cite[pp.72-76]{3}).
Let $(X,\mu )$\ be a measure space and let $\mathcal{A}_{2}(x)
\subset L^{2}(X,\mu )$\ be a closed subspace of infinite dimension. Let $%
\left\{ \Phi _{n}\right\} _{n=0}^{\infty }$ be an orthogonal basis of $%
\mathcal{A}_{2}(x) $ satisfying, for arbitrary $\xi \in X,$

\begin{equation}
\omega (\xi) :=\sum_{n=0}^{\infty }\frac{\left| \Phi _{n}(\xi) \right| ^{2}}{\rho _{n}}<+\infty ,  \label{2.1}
\end{equation}
where $\rho _{n}:=\left\| \Phi _{n}\right\| _{L^{2}(X)}^{2}$\ . Define

\begin{equation}
K(\xi ,\zeta ):=\sum_{n=0}^{\infty }\frac{\Phi _{n}(\xi)
\overline{\Phi _{n}(\zeta )}}{\rho _{n}},\text{ }\xi ,\zeta \in X.  \label{2.2}
\end{equation}
Then, $K(\xi ,\zeta )$\ is a reproducing kernel, $\mathcal{A}_{2}(x) $ is the corresponding reproducing kernel Hilbert space and $\omega
(\xi) =K(\xi ,\xi )$, $\xi \in X.$

Let $\mathcal{H}$ be another Hilbert space with $\dim \mathcal{H}=\infty $
and $\left\{ \phi _{n}\right\} _{n=0}^{\infty }$\ be an orthonormal basis of
$\mathcal{H}.$ Therefore$,$ define a coherent state as a ket vector $\mid
\xi >\in \mathcal{H}$ labeled by a point $\xi \in X$ as
\begin{equation}
\mid \xi >:=\left( \omega (\xi) \right) ^{-\frac{1}{2}%
}\sum_{n=0}^{\infty }\frac{\Phi _{n}(\xi) }{\sqrt{\rho _{n}}}%
\phi _{n}.  \label{2.3}
\end{equation}
We rewrite \eqref{2.3} using Dirac's bra-ket notation as
\begin{equation}
<q\mid \xi >=\left( \omega (\xi) \right) ^{-\frac{1}{2}%
}\sum_{n=0}^{\infty }\frac{\Phi _{n}(\xi) }{\sqrt{\rho _{n}}}%
\phi _{n}(q) .  \label{2.4}
\end{equation}
By definition, it is straightforward to show that $<\xi \mid \xi >=1$\ and
the coherent state transform $T:\mathcal{H\rightarrow A}_{2}(x)
\subset L^{2}(X,\mu )$ defined by
\begin{equation}
T\left[ \phi \right] (\xi) :=\left( \omega (\xi)
\right) ^{\frac{1}{2}}<\xi \mathsf{I}\phi >  \label{2.5}
\end{equation}
is an isometry. Thus, for $\phi ,\psi \in \mathcal{H}$, we have
\begin{equation}
<\phi \mathsf{I}\psi >_{\mathcal{H}}=<T\left[ \phi \right] \mathsf{I}T\left[
\psi \right] >_{L^{2}(x) }=\int\limits_{X}d\mu (\xi) \omega (\xi) <\phi \mid \xi ><\xi \mid \psi >  \label{2.6}
\end{equation}
and thereby we have a resolution of the identity
\begin{equation}
\mathbf{1}_{\mathcal{H}}=\int\limits_{X}d\mu (\xi) \omega
(\xi) \mid \xi ><\xi \mid ,  \label{2.7}
\end{equation}
where $\omega (\xi) $\ appears as a weight function.\\

\noindent \textbf{Remark 2.1.} The formula \eqref{2.3} can be considered as a
generalization of the canonical coherent states :
\begin{equation}
\mid \frak{z}>:=e^{-\frac{1}{2}\left| \frak{z}\right|
^{2}}\sum_{k=0}^{+\infty }\frac{\frak{z}^{k}}{\sqrt{k!}}\phi _{k},\frak{z}%
\in \Bbb{C},  \label{2.8}
\end{equation}
with $\left\{ \phi _{k}\right\} _{k=0}^{+\infty }$\ being an orthonormal
basis consisting of eigenstates of the harmonic oscillator. Here, the space $%
\mathcal{A}_{2}$ is the Bargmann space of holomorphic functions on $\Bbb{C}$
which are square integrable with respect to the Gaussian measure $e^{-\left|
\frak{z}\right| ^{2}}d\lambda \left( \frak{z}\right) $ and $\omega \left(
\frak{z}\right) \propto e^{\left| \frak{z}\right| ^{2}},\frak{z}\in \Bbb{C}$.

\section{The Barut-Girardello space}

In \cite[p.51]{1}, Barut and Girardello have considered a
countable set of Hilbert spaces $\frak{F}_{\sigma }(\Bbb{C})
,\sigma >0$ with $2\sigma =1,2,3,\cdots,$ whose elements are analytic functions
$\varphi $ on $\Bbb{C}.$ For each fixed $\sigma ,$ the inner product is
defined by
\begin{equation}
\left\langle \varphi ,\psi \right\rangle _{\sigma }:=\int\limits_{\Bbb{C}%
}\varphi (z) \overline{\psi (z) }d\mu _{\sigma
}(z) ,  \label{3.1}
\end{equation}
where
\begin{equation}
d\mu _{\sigma }(z) :=\frac{2}{\pi \Gamma (2\sigma) }%
r^{2\sigma -1}K_{\frac{1}{2}-\sigma }\left( 2r\right) rd\theta
dr,z=re^{i\theta }\in \Bbb{C},  \label{3.2}
\end{equation}
with the MacDonald function $K_{\nu }(.) $ defined by \cite[p.78]{4}: %
\begin{equation}
K_{\nu }(\xi) =\frac{1}{2}\pi \frac{I_{-\nu }(\xi)
-I_{\nu }(\xi) }{\sin \nu \pi },  \label{3.3}
\end{equation}
$I_{\nu }(.) $ denotes the modified Bessel function given by the
series
\begin{equation}
I_{\nu }(\xi) =\sum\limits_{m=0}^{+\infty }\frac{\left( \frac{1%
}{2}\xi \right) ^{\nu +2m}}{m!\Gamma \left( \nu +m+1\right) }.  \label{3.4}
\end{equation}
Precisely, $\frak{F}_{\sigma }(\Bbb{C}) $ consists of entire
functions $\varphi $ with finite norm $\left\| \varphi \right\| _{\sigma }=$
$\sqrt{\left\langle \varphi ,\varphi \right\rangle _{\sigma }}<+\infty .$
Note also that if $\varphi (z) $ is an entire function with
power series $\sum\limits_{n}c_{n}z^{n},$ then the norm in terms of the
expansion coefficients is given by
\begin{equation}
\left\| \varphi \right\| _{\sigma }=\left( \left( \Gamma (2\sigma) \right) ^{-1}\sum\limits_{n=0}^{+\infty }\left| c_{n}\right|
^{2}n!\Gamma \left( 2\sigma +n\right) \right) ^{\frac{1}{2}}.\quad \qquad
\label{3.5}
\end{equation}
Every set of coefficients $\left( c_{n}\right) $ for which the sum \eqref{3.5} converges defines an entire function $\varphi \in \frak{F}%
_{\sigma }(\Bbb{C}) .$ An orthonormal set of vectors in $\frak{F}%
_{\sigma }(\Bbb{C}) $ is given by:
\begin{equation}
\Phi _{n,\sigma }(z):=\left( \Gamma (2\sigma) \right) ^{\frac{1%
}{2}}\frac{z^{n}}{\sqrt{n!\Gamma \left( 2\sigma +n\right) }},\text{ }%
n=0,1,2,\cdots,\text{ }z\in \Bbb{C}.\qquad   \label{3.6}
\end{equation}
\quad The reproducing kernel of the Hilbert space $\frak{F}_{\sigma }(\Bbb{C}) $ can be obtained as the confluent hypergeometric limit
function $_{0}\digamma _{1}$ as
\begin{equation}
\frak{K}_{\sigma }(z,w) =\sum\limits_{n=0}^{+\infty }\frac{1}{%
(2\sigma) _{n}}\frac{\left( z\overline{w}\right) ^{n}}{n!}%
=_{0}\digamma _{1}\left( 2\sigma ;z\overline{w}\right) .  \label{3.7}
\end{equation}
Recall that (\cite[p.100]{4}):
\begin{equation}
_{0}\digamma _{1}\left( \eta ;u\right) =\sum\limits_{n=0}^{+\infty }\frac{1}{%
\left( \eta \right) _{n}}\frac{u^{n}}{n!}  \label{3.8}
\end{equation}
in which $(a) _{n}$ denotes the Pochhammer's symbol defined by $%
(a) _{0}:=1$ and
\begin{equation}
(a) _{n}:=\prod_{j=1}^{n}\left( a+j-1\right) =a\left( a+1\right)
\cdots\left( a+j-1\right) =\frac{\Gamma \left( a+n\right) }{\Gamma (a) }  \label{3.9}
\end{equation}
Making use of the relation ($\left[ 4\right] ,$ p.100):
\begin{equation}
\text{ }_{0}\digamma _{1}\left( \nu +1;-\frac{1}{4}\zeta ^{2}\right) =\Gamma
\left( \nu +1\right) \left( \frac{1}{2}\zeta \right) ^{-\nu }J_{\nu }\left(
\zeta \right) ,  \label{3.10}
\end{equation}
$J_{\nu }(.) ,\nu \in \Bbb{R}$ being the Bessel function given
by
\begin{equation}
J_{\nu }\left( \zeta \right) =\sum\limits_{m=0}^{+\infty }\frac{\left(
-1\right) ^{m}\left( \frac{1}{2}\zeta \right) ^{2m+\nu }}{m!\Gamma \left(
m+\nu +1\right) }.  \label{3.11}
\end{equation}
as well as the relation:
\begin{equation}
I_{\nu }\left( u\right) =\exp \left( -\frac{1}{2}\nu \pi i\right) J_{\nu
}\left( e^{\frac{1}{2}\pi i}u\right)   \label{3.12}
\end{equation}
for $\nu =2\sigma -1$ and $\zeta =2i|z|,$ we can write the
diagonal function $\omega _{\sigma }(z) :=\frak{K}_{\sigma
}\left( z,z\right) $ of the reproducing kernel of $\frak{F}_{\sigma }(\Bbb{C}) $ as
\begin{equation}
\omega _{\sigma }(z) =\Gamma (2\sigma) |z|^{1-2\sigma }I_{2\sigma -1}(2|z|) ,z\in
\Bbb{C}.  \label{3.13}
\end{equation}

\section{The Hardy space}

The Hardy space $\mathcal{H}\left( \prod^{+}\right) $ on the upper half of
the complex plane $\prod^{+}:=$ $\left\{ z=x+iy,x\in \Bbb{R},y>0\right\} $
consists of all functions $F(z) $ analytic on $\prod^{+}$ such
that
\begin{equation}
\sup_{y>0}\int\limits_{\Bbb{R}}\left| F\left( x+iy\right) \right|
^{2}dx<+\infty .  \label{4.1}
\end{equation}
Any function $F\left( x+iy\right) $ has a unique boundary value $f(x) $ on the real line $\Bbb{R}.$ i.e.,
\begin{equation}
\lim_{y\rightarrow 0}F\left( x+iy\right) =f(x)   \label{4.2}
\end{equation}
which is square integrable on $\Bbb{R}.$ Thus, a function $F\in \mathcal{H}%
\left( \prod^{+}\right) $uniquely determines a function $f\in $ $L^{2}(\Bbb{R}) .$ Conversely, any function $F$ can be recovered from its
boundary values on the real line by mean of the Cauchy integral \cite{5} as follows
\begin{equation}
F(z) =\frac{1}{2\pi i}\int\limits_{\Bbb{R}}\frac{f(x) }{x-z}dx,  \label{4.3}
\end{equation}
$f(x) $ being the function representing the boundary values of $%
F(z) .$ The linear space of all functions $f(x) $ is
denoted by $\mathcal{H}_{+}^{2}(\Bbb{R}) $. Since there is one
to one correspondence between functions in $\mathcal{H}_{+}^{2}\left( \Bbb{C}%
\right) $ and their boundary values in $\mathcal{H}_{+}^{2}\left( \Bbb{R}%
\right) ,$ we identify these two spaces.

Moreover, using a Paley-Wiener theorem (\cite[p.175]{6}) one can
characterize Hardy functions $f\in \mathcal{H}_{+}^{2}(\Bbb{R}) $
by the fact that their Fourier transforms
\begin{equation}
\mathcal{F}[f] (t) =\frac{1}{\sqrt{2\pi }}%
\int\limits_{\Bbb{R}}e^{-itx}f(x) dx  \label{4.4}
\end{equation}
are supported in $\Bbb{R}^{+}= [ 0,+\infty ) .$ That is,
\begin{equation}
\mathcal{H}_{+}^{2}(\Bbb{R}) =\left\{ f\in L^{2} ( \Bbb{R}) ,\mathcal{F}[f] (t) =0,\forall t<0\right\} .
\label{4.5}
\end{equation}
This last definition appear in the context of the wavelets analysis \cite{7}.

Now, since $\mathcal{F}$ is a linear isometry from $L^{2}( \Bbb{R}) $ onto $L^{2}(\Bbb{R}) $ under which the Hardy space $%
\mathcal{H}_{+}^{2}(\Bbb{R}) $ is mapped onto the space $%
L^{2}(\Bbb{R}^{+}) $ which admits the complete orthonormal
system of functions given in terms Laguerre polynomial $L_{n}^{\left( \alpha
\right) }(t) $ as
\begin{equation}
l_{n}^{\alpha }(t) :=\left( \frac{n!}{\Gamma \left( n+\alpha
+1\right) }\right) ^{\frac{1}{2}}t^{\frac{1}{2}\alpha }e^{-\frac{1}{2}%
t}L_{n}^{\left( \alpha \right) }(t) ,\quad \alpha >-1,  \label{4.6}
\end{equation}
the application of the inverse Fourier transform to the Laguerre functions
in \eqref{4.6}, $\widehat{l_{n}^{\alpha }}(x) :=\mathcal{F}^{-1}\left[ t\mapsto l_{n}^{\alpha }(t) \right] (x)$, generates a class of orthonormal rational functions
which are complete in $\mathcal{H}_{+}^{2}(\Bbb{R})$.
The obtained functions can be found in the book of J.R. Higgins \cite[p.62]{2} and are of the form:
\begin{align}
\left[ \frac{a^{1+\alpha}\Gamma(1+n+\alpha)}{n! 2\pi}\right]^{\frac 12} & 
\frac{\Gamma(1+\frac{\alpha}{2})}{\Gamma(1+\alpha)} \label{4.7}\\
\times & (ix+\frac a2)^{-(1+\alpha)}
{_{2}\digamma _{1}}\left( -n,\frac{\alpha }{2}+1,\alpha +1;\frac{2a}{2ix+a}\right) \nonumber 
\end{align}
 where $_{2}\digamma _{1}$ is the Gauss hypergeometric function defined by (\cite[p.64]{8}):
\begin{equation}
_{2}\digamma _{1}\left( a,b,c;\zeta \right) =\sum\limits_{n=0}^{+\infty }%
\frac{(a) _{n}(b) _{n}}{(c) _{n}}\frac{%
\zeta ^{n}}{n!}.  \label{4.9}
\end{equation}
For our purpose, we take $a=-1$ and we set $\alpha +1=2\sigma $ and we will be dealing with
\begin{align}
\phi _{n}^{\sigma }(x) :=& \left( \frac{\Gamma \left( \sigma +\frac{1}{2}\right) \Gamma \left( 2\sigma +n\right) }{2^{2\sigma }\sqrt{\pi }%
\Gamma (2\sigma) \Gamma (\sigma) n!}\right) ^{\frac{1}{2}}   \label{4.10}
\\ \qquad \times & \left( \frac{1}{2}-ix\right) ^{-\left( \sigma +\frac{1}{2}\right) } {_{2}\digamma _{1}}\left( -n,\sigma +\frac{1}{2},2\sigma ;\frac{1}{\frac{1}{2}-ix}\right)  \nonumber
\end{align}
as a complete orthonormal system of rational functions in the Hardy space $%
\mathcal{H}_{+}^{2}(\Bbb{R}) $.\\

\noindent \textbf{Remark 4.1. }In the particular case $\sigma =\frac{1}{2},$
the orthonormal basis $\phi _{n}^{\frac{1}{2}}(x) $ have been
discussed in \cite{9} in connection with the Hardy filter.

\section{Coherent states in the Hardy space}

Now, we combine the two basis $\left( \phi _{n}^{\sigma }(x)
\right) _{n}$ in \eqref{4.10} and $\left( \Phi _{n,\sigma
}(z)\right) _{n}$ in \eqref{3.6} according to definition \eqref{2.3} to construct for every fixed parameter $\sigma >0$\ with $%
2\sigma =1,2,3,\cdots,$\ a set of coherent states $\left( \mid z,\sigma
>\right) _{z\in \Bbb{C}}$\ labeled by points $z$\ of the complex plane $%
\Bbb{C}$\ and belonging to the Hardy space $\mathcal{H}_{+}^{2}\left( \Bbb{R}%
\right) $ as
\begin{equation}
\mid z,\sigma >:=\left( \omega _{\sigma }(z) \right) ^{-\frac{1}{%
2}}\sum_{n=0}^{\infty }\Phi _{\sigma ,n}(z)\phi _{n}^{\sigma }.  \label{5.1}
\end{equation}
with the following precisions:

\begin{itemize}
\item[$\bullet$] $ (X,\sigma )=(\Bbb{C},|z|^{2\sigma -1}K_{\frac{1}{2}%
-\sigma }(2|z|) d\lambda (z) ),$ $%
d\lambda (z)$ being the Lebesgue measure on $\Bbb{C}$

\item[$\bullet$] $ \mathcal{A}_{2}:=$ \ $\frak{F}_{\sigma }(\Bbb{C}) ,\sigma
>0$ with $2\sigma =1,2,3,\cdots$ denotes the Barut-Girardello space

\item[$\bullet$] $\omega _{\sigma }(z) =$ $\omega _{\sigma
}(z) =\Gamma (2\sigma) |z|^{1-2\sigma
}I_{2\sigma -1}(2|z|) ,z\in \Bbb{C}$ as in \eqref{3.13}.

\item[$\bullet$] $\Phi _{\sigma ,n}(z)$, $n=0,1,2,\cdots$ are the basis elements given
by \eqref{3.6}.

\item[$\bullet$] $\mathcal{H}$:=$\mathcal{H}_{+}^{2}(\Bbb{R}) $ is the
Hilbert space carrying the coherent states

\item[$\bullet$] $ \phi _{n}^{\sigma }(x) ,n=0,1,2,\cdots $ is the
orthonormal basis in \eqref{4.10}.
\end{itemize}

From \eqref{5.1}, the coherent states $\mid z,\sigma >$ are defined
by their wave functions through the series expansion
\begin{equation}
<x\mid \sigma ,z>=\left( \omega _{\sigma }(z) \right) ^{-\frac{1%
}{2}}\sum_{n=0}^{\infty }\frac{z^{n}}{\sqrt{(2\sigma)
_{n}\Gamma \left( n+1\right) }}\phi _{n}^{\sigma }(x)   \label{5.2}
\end{equation}
Explicitly, we have that
\begin{eqnarray}
&<&x\mid \sigma ,z>=\left( \Gamma (2\sigma) |z|^{1-2\sigma }I_{2\sigma -1}(2|z|) \right) ^{-\frac{1%
}{2}}  \label{5.3} \\
&&\times \left( \frac{1}{2}-ix\right) ^{-\left( \sigma +\frac{1}{2}\right)
}\sum_{n=0}^{\infty }\frac{z^{n}}{\sqrt{(2\sigma) _{n}\Gamma
\left( n+1\right) }}  \nonumber \\
&&\times \left( \frac{\Gamma \left( \sigma +\frac{1}{2}\right) \Gamma \left(
2\sigma +n\right) }{2^{2\sigma }\sqrt{\pi }\Gamma (2\sigma)
\Gamma (\sigma) n!}\right) ^{\frac{1}{2}} {_{2}\digamma
_{1}}\left( -n,\sigma +\frac{1}{2},2\sigma ;\frac{1}{\frac{1}{2}-ix}\right)
\nonumber
\end{eqnarray}
We write the Gauss hypergeometric function $_{2}\digamma _{1}$ in terms of
the Meixner polynomial (\cite[p.346]{8}) as
\begin{equation}
_{2}\digamma _{1}\left( -n,-u,b;1-\frac{1}{c}\right) =M_{n}\left(
u,b;c\right)   \label{5.4}
\end{equation}
Next ext we make use of the relation (\cite[p.349]{8})
\begin{equation}
\sum\limits_{n=0}^{+\infty }M_{n}\left( u,b;c\right) \frac{\zeta ^{n}}{n!}%
=e^{\zeta }{_{1}\digamma _{1}}\left( -u,b,\left( \frac{1-c}{c}\right)
\zeta \right)   \label{5.5}
\end{equation}
where for $_{1}\digamma _{1}(.) $ is the Kummer's function (\cite[p.262]{10})
\[
_{1}\digamma _{1}\left( a,\beta ;\frak{z}\right) :=\frac{\Gamma \left( \beta
\right) }{\Gamma (a) }\sum\limits_{j=0}^{+\infty }\frac{\Gamma
\left( \beta +j\right) }{\Gamma \left( a+j\right) }\frac{\frak{z}^{j}}{j!}
\]
and
\begin{equation}
\zeta =z,u=-\left( \sigma +\frac{1}{2}\right) ,b=2\sigma ,c=-\frac{\frac{1}{2%
}-ix}{\frac{1}{2}+ix}  \label{5.6}
\end{equation}
Therefore, we get that
\begin{equation}
\sum\limits_{n=0}^{+\infty }\frac{z^{n}}{n!}\text{ }_{2}\digamma _{1}\left(
-n,\sigma +\frac{1}{2},2\sigma ;\frac{1}{\frac{1}{2}-ix}\right) =e^{z}\text{
}_{1}\digamma _{1}\left( \sigma +\frac{1}{2},2\sigma ,\frac{-z}{\frac{1}{2}%
-ix}\right)   \label{5.7}
\end{equation}
and the wave functions of these coherent states
\begin{eqnarray}
&<&x\mid z,\sigma >=\left( \frac{\Gamma \left( \sigma +\frac{1}{2}\right) }{%
2^{2\sigma }\sqrt{\pi }\Gamma (\sigma) }\right) ^{\frac{1}{2}%
}\left( \Gamma (2\sigma) |z|^{1-2\sigma
}I_{2\sigma -1}(2|z|) \right) ^{-\frac{1}{2}}e^{z}
\label{5.8} \\
&&\times \left( \frac{1}{2}-ix\right) ^{-\left( \sigma +\frac{1}{2}\right)
}{}_{1}\digamma _{1}\left( \sigma +\frac{1}{2},2\sigma ,\frac{-z}{\frac{1}{2}%
-ix}\right)   \nonumber
\end{eqnarray}

Now, according to \eqref{2.5} the coherent state transform
corresponding to these coherent states is the isometry mapping the Hilbert
space $\mathcal{H}_{+}^{2}(\Bbb{R}) $ into the weighted Bergman
space $\frak{F}_{\sigma }(\Bbb{C}) $ as
\begin{equation}
T_{\sigma }:\mathcal{H}_{+}^{2}(\Bbb{R}) \rightarrow \frak{F}%
_{\sigma }(\Bbb{C})   \label{5.9}
\end{equation}
defined, according to \eqref{4.1} by
\begin{equation}
T_{\sigma }[f] (z) :=\left( \omega _{\sigma }(z) \right) ^{\frac{1}{2}}<z,\sigma \mid f>  \label{5.10}
\end{equation}
Explicitly,
\begin{equation}
T_{\sigma }[f] (z) =\int\limits_{\Bbb{R}}\mathcal{K}%
_{\sigma }(z,x) \overline{f(x)}dx,f\in \mathcal{H}_{+}^{2}(\Bbb{R}) ,z\in \Bbb{C}  \label{5.11}
\end{equation}
with the kernel function
\begin{equation}
\mathcal{K}_{\sigma }(z,x) :=\frac{1}{2^{\sigma }\pi ^{\frac{1}{4%
}}}\left( \frac{\Gamma \left( \sigma +\frac{1}{2}\right) }{\Gamma (\sigma) }\right) ^{\frac{1}{2}}\left( \frac{1}{2}-ix\right) ^{-\left(
\sigma +\frac{1}{2}\right) }e^{z}\text{ }_{1}\digamma _{1}\left( \sigma +%
\frac{1}{2},2\sigma ,\frac{-z}{\frac{1}{2}-ix}\right)   \label{5.12}
\end{equation}

Finally, recalling \eqref{2.7}  one can write that the coherent
states $\mid z,\sigma >$\ labeled by points $z\in \Bbb{C}$\ solve the
identity of the Hardy space $\mathcal{H}_{+}^{2}(\Bbb{R}) $ as
\begin{equation}
1_{\mathcal{H}_{+}^{2}(\Bbb{R}) }=\int\limits_{\Bbb{C}}d\lambda
(z) \omega _{\sigma }(z) \mid z,\sigma ><z,\sigma
\mid .  \label{5.13}
\end{equation}

\end{document}